\documentstyle[12pt]{article}
\begin{document}
\renewcommand{\thefootnote}{\fnsymbol{footnote}}
\baselineskip = 24pt

\title{ Bhabha scattering at LEP1. Calorimeter event selection }
\vspace{1.5cm}
\author {G.I.Gakh, V.Yu.Gonchar, N.P.Merenkov}
\vspace{5mm}
\date{}
\maketitle
\begin{center}
{\it National Scientific Centre
"Kharkov Institute of Physics and Technology" \\ }
{\it Akademicheskaya 1, 310108, Kharkov, Ukraine \\ }
\end{center}
\vspace{1cm}
\begin{center}
{Abstract}
\end{center}
{\small
{Analytical calculations have been used to describe the calorimeter
event selection in small--angle electron--positron scattering at LEP1. 
The first--order QED correction to the Born cross--section has been 
derived with leading log and next--to--leading log approximations. The 
second-- and third--order corrections are computed with leading accuracy. 
Our analytical results are illustrated by Tables and compared with 
the corresponding results obtained with the of Monte Carlo generator 
BHLUMI.\\ }}

\vspace{2cm}

PACS \ \
12.15.Lk, \ \ 12.20.-m, 12.20.Ds, 13.40.-f

\newpage
\large
\baselineskip = 24pt

\begin{center}
\section{Introduction}
\end{center}
\vspace{0.05cm}

The test of the Standard Model (SM) based on the unified theory of
electroweak interaction and quantum chromodynamics, is one of the
primary aims of experiments at LEP1 and LEP2. The small--angle
Bhabha scattering (SABH) process is used to measure the luminosity of the
corresponding electron--positron collider. Being the normalization factor,
the cross--section for the SABH process affects all observable
cross--sections and is a significant component in both the
precision measurements of the SM parameters and
investigation of new physics at LEP1 [1].

By now a purely experimental precision better than 0.1 percent for the 
luminosity determination has been achieved at LEP1 
[2].  However, to obtain the total value a systematic theoretical error 
must be added. Ths latter is determined by the accuracy of 
theoretical description of SABH cross--section, that takes into 
consideration the specidic features of event selection and the detector 
geometry at LEP1.  Therefore, there is a need to decrease its precision to 
the level of experimental one or better it.  It is no mere chance that 
lately much attention has been given to the theoretical investigation of 
the SABH process [3--14].

The theoretical calculation of the SABH cross--section at LEP1 involves two
somewhat different tasks. The first one is to privide an adequate 
description of the experimental restrictions imposed on the event 
selection in terms of final--particle phase space. The second task 
consists in writing the matrix element squared within the required 
accuracy. There are two methods for theoretical investigation of SABH at 
LEP1: the method based on the Monte Carlo event generator [3--6] and 
the method resting on semianalytical calculations [7--14].

The advantage of the Monte Carlo method is that it enables simulation
of detectors and event selection schemes of different types. The modern
Monte Carlo generators employ some additional procedures
(Yennie--Frautchi--Suura exponentiation [15], electron structure function
method [16] and effective--coupling--constant approach [17]) to eliminate
problems associated with the infrared divergence, and to take into account 
the leading corrections in the higher orders of the perturbation theory.  
Although a part of the second--order
next--to--leading correction is compensated due to the exponential form of
the electron structure function (see, for example,  [7]), the total
second--order correction, including $\alpha^2\ln(Q^2/m^2)$ terms
(here $Q^2$ is the typical value of the transfer momentum squared), remains
undeterminated.

The advantage of the analytical method is that it allows the use
of the exact matrix element squared, based on essential Feynman diagrams. 
The infrared problem in the context of this approach can be solved in the 
usual way by taking into account virtual, real (soft and hard) photon 
emission and pair production in every order of the perturbation theory. 
The shortcoming of the analytical approach is its low mobility with 
respect to the change in the experimental conditions used for the event 
selection. Nevertheless, the analytical calculations are very important 
because they privide checking many Monte Carlo calculations for different 
"ideal" detectors.

At present, four types of "ideal" event selections are used
for the comparison of various MC -- generators. One of them is completely
inclusive as to the final electron and positron. In this case, the event is
the simultaneous detection of the scattered electron and positron with
the invariant mass higher than a certain threshold value. The ring--shaped
symmetric or asymmetric detectors are used for the detection. this
method of event selection is called in Ref.[3] as BARE1.

The other three methods of event selection, called in Ref.[3] as CALO1,
CALO2 and SICAL2, are calorimetric. If the final particles (photons,
electrons and positrons) move almost in parallel to each other, i.e., they
form a cluster, then the total energy of all particles in the
cluster is determined at the calorimeter event selection. If the final
particles do not form a cluster, then CALO1 and CALO2 detect only the
scattered electron and positron, whereas SICAL2 does not discriminate
photons and electrons (or positrons). CALO1 and CALO2 differ in the
shape and size of the cluster. The detectors of the same geometry as
in the case of BARE1 are used particle detection.

It is most natural to apply the analytical approach to the same
methods of the event selection, which are used in the MC--generators. The
analytical formulae for the SABH cross--section at LEP1 were published for
the BARE1 method of event selection in the cases of symmetric [11--13] 
and asymmetric [14] detectors. Those formulae include the full 
first--order correction as well as the second--order correction with 
leading and next--to--leading accuracy.  They also include the 
third-- order correction in the leading approximation. It is just these 
contributions that
must be calculated in order to achieve the required accuracy. Some brief 
analytical results for the CALO1 and CALO2 event selections in the case of 
symmetric detectors have been given in Ref.[14,25].

In this paper we present full analytical calculations of the SABH
cross--section at LEP1 for CALO1 and CALO2 in the most general
case of asymmetric detectors. The corresponding formulae include the
same contributions as for the BARE1 method, except the second--order
next--to--leading contribution. As to the SICAL2 method, we hope to
investigate it in the following publications.

The paper is organized as follows. In Section 2 we introduce the
"observable" cross--section $\sigma_{obs}$ with allowance for the
experimental restrictions on the angles and energies of the detected
particles. We also calculate the first--order correction for the BARE1
method in the case of asymmetric detectors. The formulae obtained are
widely used below. In Section 3 the first--order correction is considered
for the CALO1 method, and in Section 4 -- for the CALO2. In Section 5
we present the formulae for the leading contributions in the second and
third orders using the electron structure function method. These formulae
are universal because they are independent of the shape and size of the
cluster.  In Section 6 we present the Tables that illustrate the
comparison of our analytical results with the corresponding calculations
made with the MC--generator BHLUMI [3].  \vspace{0.3cm}
\begin{center}
\section{First--order correction for BARE1}
\end{center}
\vspace{0.05cm}

For the investigation of radiative corrections to the Born
cross--section of SABH at LEP1 it is convenient to introduce the
dimensionless
quantity
\begin{equation}\label{1}
\Sigma = \frac{1}{4\pi\alpha^2}Q_1^2\sigma_{obs}  ,
\end{equation}
where $Q_1^2 = \varepsilon^2\theta_1^2$ ($\varepsilon $ is the
beam energy, and $ \theta_1$ is the minimal angle of the
wide detector). The "observed" cross--section $\sigma_{obs}$
is determined by the following formula in the BARE1 case
\begin{equation}\label{2}
\sigma_{obs} = \int
dx_1dx_2\Theta d^2q_1^{\bot}d^2q_2^{\bot}\Theta_1^c
\Theta_2^c\frac{d\sigma (e^{+}+e^{-} \rightarrow e^{+}+e^{-}+X)}{dx_1dx_2
d^2q_1^{\bot}d^2q_2^{\bot}} \ ,
\end{equation}
where X denotes undetected final particles, and $x_1\ (x_2)$ and
$ \vec q_1^{\bot} \ (\vec q_2^{\bot})$ are the energy fraction and the
transverse component of the momentum of the electron (positron) in the
final state. The functions
$\Theta_i^c \  (i=1,2)$ take into account the angular cuts, and the
function $\Theta $ takes into account the cutoff on the energies of
the detected electron and positron
$$ \Theta_1^c = \theta(\theta_3 -
\theta_{-})\theta(\theta_{-} - \theta_1),\ \ \Theta_2^c = \theta(\theta_4
- \theta_{+})\theta(\theta_{+} - \theta_2),\ \ \Theta = \theta
(x_1x_2-x_c), $$
\begin{equation}\label{3} \theta_{-} = \frac{\mid\vec
q_1^{\bot}\mid}{x_1\varepsilon},\ \ \ \theta_{+} = \frac{\mid\vec
q_2^{\bot}\mid}{x_2\varepsilon} \ .
\end{equation}

In the case of asymmetrical ring--shaped detectors we have
$$\theta_3 > \theta_4 > \theta_2 >
\theta_1,\ \ \rho_i =\frac{\theta_i}{\theta_1} > 1  \ .$$
Below we shall assume, for definiteness, that the final electron is
registered by wide detector $(\theta_3, \theta_1),$ and positron --
by narrow one $(\theta_4, \theta_2).$

The Born contribution to $\Sigma$, for SABH process at LEP1, is determined
as follows
\begin{equation}\label{4}
\Sigma_B = \int\limits_{\ \ \ \rho_2^2}^{\rho_4^2}\frac{dz}{z^2}(1-
\frac{z}{2} \theta_1^2) \ ,
\end{equation}
and in symmetrical case the limits of integration are $(\rho_3^2 , 1).$
The Born cross--section is the same both for the inclusive and calorimeter
event selections. Formula (4) includes the contribution of the scattering
diagram and its interference with the annihilation one. The contribution
of the annihilation diagram is proportional to $\theta_1^4$ and may be
omitted within the scope of  0.1\% .
When calculating the radiative corrections to the cross--section (4), we
systematically omit the terms proportional to $\theta_1^2.$

The first--order correction to the cross--section of
electron--positron scattering includes the contributions connected with 
radiation of the virtual and real (soft and hard) photons and vacuum 
polarization due to the light fermions including hadrons. The contribution
of vacuum polarization is universal, and therefore we shall not take it
into account below.  \begin{equation}\label{5} \Sigma_1 = \Sigma_{V+S} +
\Sigma_H + \Sigma^H \ .  \end{equation} The correction due to the virtual
and soft real photon emission (with the energy less than
$\Delta\varepsilon ,\Delta\ll1 $)  can be written as follows [12] (in this
case $ x_1 = x_2 = 1,\ \ \vec q_1^{\bot} + \vec q_2^{\bot} = 0 $)
\begin{equation}\label{6} \Sigma_{V+S} = 2\frac{\alpha}{\pi}\int
\limits_{\ \ \ \rho_2^2}^{\rho_4^2} \frac{dz}{z^2}[2(L-1)ln\Delta
+\frac{3}{2}L -2] ,\ \   L = \ln\frac{\epsilon^2 \theta_1^2z}{m^2} \ ,
\end{equation}
where $ z= \vec q_2^{\bot 2}/Q_1^2 $,\ $m$ is the electron mass.

The second term on the right side of Eq.(5) is due to the emission
of the hard( with energy greater than $\Delta\varepsilon$) photon by the
positron registered by the narrow detector. In this case we have
\begin{equation}\label{7}
X = \gamma(\vec k^{\bot}, 1-x_2), \ \ x_1=1 \ ,\
\vec k^{\bot} + \vec q_1^{\bot} + \vec q_2^{\bot} = 0,\ \ \ x_c
<x_2<1-\Delta \ ,
\end{equation}
where $\vec k^{\bot}$ is the transverse component of the three--momentum
of the hard photon. The corresponding contribution can be obtained by the
 integration of the bremsstrahlung differential cross--section in the
electron--positron collision over the region
\begin{equation}\label{8}
1 < z < \rho_3^2 ,\ \
x^2\rho_2^2 < z_1 = \frac{\vec q_1^{\bot 2}}{Q_1^2} < x^2\rho_4^2 ,\ \ -1
< cos\varphi < 1 \ ,
\end{equation}
where $\varphi$ is the angle between two--dimensional vectors
$\vec q_1^{\bot}$ and $\vec q_2^ {\bot}.$ The result is defined
by the formula
\begin{equation}\label{9}
\Sigma_{H} = \frac{\alpha}{2\pi}\int\limits_{\ \ \
 1}^{\rho_3^2}\frac{dz}{z^2}\int\limits_{\ \ \ x_c}^
{1-\Delta}\frac{1+x^2}{1-x}dx\biggl[(L-1)(\Delta_{42}+\Delta_{42}^{(x)}) +
{K}(x,z;\rho_4,\rho_2)\biggr] \ ,
\end{equation}
where
$$ \widetilde{K} = \frac{(1-x)^2}{1+x^2}(\Delta_{42}+\Delta_{42}^{(x)}) +
\Delta_{42}\widetilde L_1 + \Delta_{42}^{(x)}\widetilde L_2 + (\overline
\theta_4^{(x)} - \theta_2^{(x)})\widetilde L_3+ (\overline\theta_4 - \theta_2)
\widetilde L_4 \ , $$
$$ \widetilde L_1 = \ln\left|{(z-\rho_2^2)(\rho_4^2-z)x^2} \over
{(x\rho_4^2-z) (x\rho_2^2-z)}\right| \ ,\ \  \widetilde L_2
=\ln\left|{(z-x^2\rho_2^2)(x^2\rho_4^2-z)} \over
{x^2(x\rho_4^2-z)(x\rho_2^2-z)}\right| \ ,$$
\begin{equation}\label{10}
\widetilde L_3 = \ln\left|{(z-x^2\rho_2^2)(x\rho_4^2-z)} \over
{(x^2\rho_4^2-z) (x\rho_2^2-z)}\right|,\ \ \ \widetilde L_4
=\ln\left|{(z-\rho_2^2)(x\rho_4^2-z)} \over
{(\rho_4^2-z)(x\rho_2^2-z)}\right| \ ,
 \end{equation}
and the following designations are used for $\theta$ -- functions:
$$ \Delta_{42}^{(x)} = \theta_4^{(x)} - \theta_2^{(x)},\ \ \ \Delta_{42} =
\theta_4 - \theta_2 \ ,$$
\begin{equation}\label{11} \theta_{i}^{(x)} =
\theta(x^2\rho_{i}^2-z),\ \ \theta_{i} = \theta(\rho_{i}^2 -z),\ \
\overline\theta{i}^{(x)} = 1 - \theta_{i}^{(x)},\ \ \ \overline\theta_{i}
= 1 - \theta_{i} \ .
\end{equation}

The term $\Sigma^{H}$, related to the emission of the hard photon
by electron registered by means of the wide detector, can be obtained
from (9) by change of the integration limits over $z$:
$(\rho_4^2, \rho_2^2)$ instead of $(\rho_3^2, 1)$ and
substitution
\begin{equation}\label{12} \Delta_{42}
\rightarrow 1\ , \ \ \Delta_{42}^{(x)} \rightarrow \theta_3^{(x)}\ ,\ \ \
K(x,z;\rho_4,\rho_2) \rightarrow K(x,z;\rho_3,1)
\end{equation}
in the integrand. It is easy to see that the contribution, including the
analogue of $\widetilde L_4 \ ,$ vanishes, and it is caused by
the specific character of the accompanying $\theta$ -- functions.

The individual parts on the right side of Eq.(5) depend on the auxiliary
parameter $\Delta$, but this dependence disappeares in the sum, so that
the first--order correction can be written as follows
$$ \Sigma_1 = \frac{\alpha}{2\pi}\biggl\{\int\limits_{\ \ 1}^{\rho_3^2}
\frac{dz}{z^2}
\int\limits_{\ \ \ x_c}^{1}\biggl[(-\frac{1}{2}\delta(1-x) + (L-1)P_1(x))
(\Delta_{42}+\Delta_{42}^{(x)}) +
\frac{1+x^2}{1-x}K(x,z;\rho_4,\rho_2)\biggr]dx $$
\begin{equation}\label{13}
+ \int\limits_{\ \
\rho_2^2}^{\rho_4^2}\frac{dz}{z^2} \int\limits_{\ \ \
x_c}^{1}\biggl[(-\frac{1}{2}\delta(1-x) + (L-1)P_1(x))(1 + \theta_3^{(x)}) +
\frac{1+x^2} {1-x}K(x,z;\rho_3,1)\biggr]dx\biggr\} \ ,
\end{equation}
where $P_1(x)$ determines the iterative form of the nonsinglet electron
structure function (see, for examàle, Ref.[7]) $$ P_1(x) =
\frac{1+x^2}{1-x}\theta(1-x-\Delta) + (2ln\Delta + \frac{3}
{2})\delta(1-x) ,\ \ \Delta \rightarrow 0 \ .$$

In order to make the cancellation of the $\Delta$ -- dependence on the
right side of Eq.(10) more transparent, one may use the following
relations
$$\int\limits_{\ \ \ x_c}^{1}P_1(x)dx = - \int\limits_{\ \ \
0}^{x_c}\frac{1+x^2}{1-x}dx \ , \ \ \ \int\limits_{\ \ \
x_c}^{1}P_1(x)\overline\theta_3^{(x)}dx = \overline\theta_3^
{(x_c)}\int\limits_{\ \ \ x_c}^{\sqrt{z}/\rho_3}\frac{1+x^2}{1-x}dx \ , $$
$$\int\limits_{\ \ \ x_c}^{1}P_1(x)\overline\Delta_{42}^{(x)}dx = \theta_4
\overline\theta_4^{(x_c)} \int\limits_{\ \ \
x_c}^{\sqrt{z}/\rho_4}\frac{1+x^2}{1-x}dx - \theta_2 \overline
\theta_2^{(x_c)}\int\limits_{\ \ \
x_c}^{\sqrt{z}/\rho_2}\frac{1+x^2}{1-x}dx \ ,$$
$$\overline\Delta_{42}^{(x)} = \Delta_{42} - \Delta_{42}^{(x)} \ .$$

As we already noted, the right side of formula (13) defines the
first--order correction to the Born cross--section of SABH at LEP1
with switched off vacuum polarization. The last one can be included by
inserting the quantity $[1-\Pi(zQ_1^2)] ^{-2}$ in the integrand (as to
$\Pi$ see Ref. [3] and the bibliography cited there).

The upper line on the right side of Eq.(13) corresponds to the emission of
the real and virtual photons by a positron and lower one -- by an
electron. Besides, the terms that are accompanied by the
$x$--depending ($x$--independing) $\theta$ -- functions describe
the emission in the initial (final) state. This information is
very important for the investigation of calorimeter event
selection.

But before to move on this investigation , we note one more
circumstance. When deriving the Eq.(13), we systematically ignore the
terms of the order of $\theta^2\approx
Q^2/S \ (S = 4\varepsilon)$ in comparison with a unity. However, as it is
known [18], the terms of this type have a double--logarithmic asymptotics.
Their contribution to $\Sigma_1$ equals parametrically to [12,19]
$$\frac{\alpha}{\pi}\frac{Q^2}{S}\ln^2\frac{Q^2}{S},$$ and it amounts to
about $10^{-4}$ in the LEP1 conditions. Therefore, we expect the systematic
deviation of our calculations in the first order of the
perturbation theory from the corresponding results of MC--generator
BHLUMI, which completely includes such contributions on the 0.01\%
level (see Section 6).

\vspace{0.3cm}
\begin{center}
\section{Cone--shaped cluster CALO1}
\end{center}
\vspace{0.05cm}

In the case of calorimeter event selection the detector does not
discriminate the particles moving near the same direction, i.e. producing
a cluster. In the first order of perturbation theory the cluster can be
formed only by two particles: photon and electron (or photon and
positron). For the definiteness we shall talk about the positron cluster,
although the same will be refer to the electron one.

If a photon and positron belong to a cluster $({\bf \gamma, e^+)
\in CL},$ the detector measures the total energy of the cluster, and its
position is determined by the positron position. If a photon and positron
do not form a cluster $({\bf \gamma, e^+) \not\in CL},$ the event looks
exactly the same as in the case of BARE1, i.e. there exists the detection
threshold on the positron energy and the photon is entirely not detected.
The cutoff on the positron energy can be written symbolically in the form
\begin{equation}\label{14}
\int\limits_{\ \ \ x_c}^{1}dx + \int\limits_{\ \ 0}^{x_c}
[(\gamma, e^+) \in CL]dx \ \ \equiv \int\limits_{\ \ 0}^{1}dx - \int\limits_
{\ \ 0}^{x_c}[(\gamma, e^+) \not\in CL]dx \ .
\end{equation}
It is convenient to use the left side of Eq.(14) in order to take into
account the initial--state emission and right side of Eq.(14) for the
description of the final--state one. As follows from Eq.(14), in the
calorimeter event selection the correction to the cross--section can be
represented in the form of two--term sum.  One of these terms is a
universal, i.e.,  it does not depend on the cluster specific and another
one is determined by its form and size  \begin{equation}\label{15}
\Sigma_1 = \Sigma_1^{^{un}} \ \ + \ \ \Sigma_1^{^{cl}} \ .  \end{equation}

The contribution of the universal part $\Sigma_1^{^{un}}$ can be obtained
with the help of the formula (13). For this one must retain without
change the correction caused by the initial--state emission and
integrate the correction, caused by the final--state emission, for $x$
going from 0 to 1. The result has the following form $$ \Sigma_1^{^{un}} =
\frac{\alpha}{2\pi}\int\limits_{\ \ \ 1}^{\rho_3^2}
\frac{dz}{z^2}\biggl\{-\Delta_{42}^{(x)} + \int\limits_{\ \ \ x_c}^{1}
[((L-1)P_1(x) + 1 - x + \frac{1+x^2}{1-x}\widetilde L_2)\Delta_{42}^{(x)}
+(\overline\theta_4^{)x)}-\theta_2^{(x)})\widetilde L_3]dx +$$
$$+\int\limits_{\ \ \ 0}^{1}[(1-x + \frac{1+x^2}{1-x}\widetilde
L_1)\Delta_{42} + \frac{1+x^2}{1-x}(\overline\theta_4-\theta_2)\widetilde
L_4]dx\biggr\} + $$
$$ + \frac{\alpha}{2\pi}\int\limits_{\ \ \ \rho_2^2}^{\rho_4^2}\frac{dz}{z^2}
\biggl\{-1 + \int\limits_{\ \ \ x_c}^{1}[((L-1)P_1(x) + 1-x
+\frac{1+x^2}{1-x} L_2)\theta_3^{(x)} + \overline\theta_3^{(x)}L_3]dx + $$
\begin{equation}\label{16}
 + \int\limits_{\ \ \ 0}^{1} (1-x + \frac{1+x^2}{1-x}L_1)dx\biggr\} ,
\end{equation}
where $L_i$ are obtained from $\widetilde L_i$ by the substitution
$\rho_3$ instead of $\rho _4$ and 1 instead of $\rho_2.$ In the case of
symmetrical ring--shaped detectors it is necessary to put
$\rho_3 = \rho_4, \ \rho_2 = 1$ on the right side of Eq.(16).

We see that owing to the well--known relation
$$ \int\limits_{\ \ \ 0}^{1}P_1(x) = 0 \ , $$
the final--state emission does not give rise to large logarithm $L$ in the
cross--section in accordance with the Lee--Nauenberg theorem about the
cancellation of the mass singularities when summing over all possible
states [21].

In order to obtain the part of the correction, determined by the shape and
size of the cluster, one may use the simplified form of the bremsstrahlung
differential cross--section, that describes the so--called semicollinear
kinematics (see, for example, Refs. [14,22]). In the case of single--photon
emission it simply means, that one can neglect the electron mass
everywhere. The corresponding expression for the dimensionless quantity
$\Sigma,$ related to the photon emission by positron, is determined as
follows \begin{equation}\label{17} d\Sigma = \frac{\alpha d\varphi
dzdz_1(1+x^2)}{4\pi Q_1^2z(z_1-xz)}\bigl[
\frac{1}{z_1+z+2\sqrt{z_1z}cos\varphi} - \frac{x}{z_1+x^2z+2x\sqrt{z_1z}
cos\varphi}\bigr]dx .
\end{equation}

The cutoff on the $ z, z_1$ and $ \varphi $ variables depends, of course,
on the cluster shape. The CALO1 cluster is the cone with the cone
semi-angle $\delta = 0.01$ around the final--positron 3--momentum. In
order to include the initial--state emission it is necessary to have
(according to Eq. (14)) the cutoff on these variables when photon and
positron belong to a cluster
\begin{equation}\label{18}
\mid\sqrt{z_1} - x\sqrt{z}\mid <
x(1-x)\lambda ,\ \  -1  < cos\varphi  < -1 + \frac{\lambda^2x^2(1-x)^2 -
(\sqrt{z_1} - x\sqrt{z})^2}{2x\sqrt{zz_1}},\ \ \lambda =
\frac{\delta}{\theta_1}
\end{equation}

For the calculation of the correction caused by the final--state emission
it is necessary to use the cutoff when photon and positron do not form a
cluster
$$ -1 <\cos\varphi < 1,\ \ \ |\sqrt{z_1} - x\sqrt{z}| > \lambda
x(1-x);$$
\begin{equation}\label{19} 1 > \cos\varphi > -1 +
     \frac{\lambda^2x^2(1-x)^2 - (\sqrt{z_1}-x\sqrt
{z})^2}{2x\sqrt{zz_1}},\ \ |\sqrt{z_1} - x\sqrt{z}| < \lambda x(1-x).
\end{equation}

We pay attention to the fact, that formally the cuts (18) and (19)
determine also the collinear and semicollinear kinematic regions,
respectively (in the case of the final--state emission [12,13,22]). The
only, but very essential difference is that, in the considered case , the
parameter $\lambda$ is of the order of unity, whereas when determining the
kinematic regions, it is required that this parameter was much less than
unity.

Integration of Eq.(17), as well as a similar formula including photon
emission by an electron, over regions (18) and (19) leads to the following
contribution to the first--order correction
\begin{equation}\label{20} \Sigma_1^{^{cl}} = \Sigma_{1i}^{^{cl}} +
     \Sigma_{1f}^{^{cl}}\ , \end{equation}
     $$\Sigma_{1i}^{^{cl}} =
\frac{\alpha}{2\pi}\int\limits_{\ \ \ 0}^{x_c}\frac{1+x^2}
{1-x}dx\int\frac{dz}{z^2}\int dz_1(\Psi +
\widetilde\Psi)\Phi(z_1,z;\lambda,x), $$
$$\Sigma_{1f}^{^{cl}} =
\frac{\alpha}{2\pi}\int\limits_{\ \ \ 0}^{x_c}\frac{1+x^2}
{1-x}dx\biggl\{\int\limits_{\ \ \ a_0^2}^{b^2} \frac{dz} {z^2}\biggl(\ln
\left|\frac{x\rho_3^2-z}{\rho_3^2-z}\right| + l_+\biggr) + \int\limits_{\
\ \ a^2}^{b_0^2} \frac{dz}{z^2}\biggl(\ln \left|\frac{x-z} {1-z}\right| +
l_-\biggr) +$$
$$+ \int\limits_{\ \ \
1}^{\rho_3^2}\frac{dz}{z^2}[\theta(\tilde a_0^2-z) - \theta (z-\tilde
b_0^2)]\widetilde L_4 + \int\limits_{\ \ \ \tilde a_0^2}^{\tilde
b^2}\frac{dz} {z^2}\biggl(\ln \left|\frac{x\rho_4^2-z}{\rho_4^2-z}\right|
+ l_+\biggr) + \int\limits_{\ \ \ \tilde a^2}^{\tilde
b_0^2}\frac{dz}{z^2}\biggl(\ln \left|\frac{x\rho_2^2-z}{\rho_2^2-z}\right|
+ l_-\biggr)  + $$
\begin{equation}\label{21} +\int\frac{dz}{z^2}\int dz_1
(\Psi + \widetilde\Psi)F(z_1,z;\lambda,x)\biggr\}, \  \ l_{\pm} =
\ln\frac{\lambda[2\sqrt{z}\pm\lambda(1-x)]}{z\mp2x\lambda\sqrt{z}
-\lambda^2x(1-x)},
\end{equation}
where the integration limits over $z$ (in the squared brackets) and
$z_1$ (in the round brackets) are defined with the help of
$\Psi$ and¨ $\widetilde\Psi$ quantities
 $$\Psi = [a^2,a_0^2](x^2z_+,x^2) + [b^2,a^2](x^2z_+,x^2z_-) +
[b_0^2, b^2] (x^2\rho_3^2,x^2z_-),\ \ z_{\pm} = (\sqrt{z}\pm
\lambda(1-x))^2 \ , $$
$$\widetilde\Psi = [\tilde a^2,\tilde
a_0^2](x^2z_+,x^2\rho_2^2) + [\tilde b^2,\tilde a^2](x^2z_+,x^2z_-) +
[\tilde b_0^2, \tilde b^2] (x^2\rho_4^2,x^2z_-) \ , $$ $$z_{\pm} =
(\sqrt{z} \pm \lambda(1-x))^2\ , \ \ a_0 = \rho_2\ , \ \ b_0 = \rho_4, \ \
a = max(\rho_2,1+\lambda(1-x))\ , $$
$$ b = min(\rho_4\ , \rho_3 -
\lambda(1-x))\ , \ \ \tilde a_0 = max(1,\rho_2-\lambda(1-x))\ , \ \ \tilde
a = \rho_2+\lambda(1-x)\ , $$
\begin{equation}\label{22} \tilde b =
\rho_4-\lambda(1-x)\ , \ \ \tilde b_0 = min(\rho_4+\lambda(1-x)\ ,
\rho_3).
\end{equation}
The $\Phi$ and $F$ functions beloning to the right side of Eq.(21) are
determined as follows:
$$ \Phi = \frac{2}{\pi}\biggl(\frac{1}{z_1-xz} + \frac{1}{z-z_1}\bigg)
\arctan\bigg\{\frac{z-z_1}{(\sqrt{z}-\sqrt{z_1})^2}R\biggr\}\ ,$$
$$ F = \frac{2}{\pi}\biggl(\frac{1}{z_1-xz} - \frac{1}{z_1-x^2z}\bigg)
\arctan\bigg\{\frac{(\sqrt{z_1}-x\sqrt{z})^2}{z_1-x^2z}R^{^{-1}}\biggr\}\ , $$
\begin{equation}\label{23}
R = \sqrt{\frac{\lambda^2x^2(1-x)^2 - (\sqrt{z_1}-x\sqrt{z})^2}
{(\sqrt{z_1}+x\sqrt{z})^2 - \lambda^2x^2(1-x)^2}} \ .
\end{equation}

The formula (21) is valid in the general case of the asymmetrical detectors .
In the symmetrical case of the wide detectors it is necessary to put
$\rho_4 = \rho_3$ and $\rho_2 = 1$ in Eq.(21)
\vspace{0.3cm}
\begin{center}
\section{Pyramidal cluster CALO2}
\end{center}
\vspace{0.05cm}

The cluster CALO2 has a pyramidal form. Its section by a plane
perpendicular to the beam axis represents a rectangle in the angular
coordinates $(\theta,\varphi),$ and the centre of it is determined by the
position of the positron 3--momentum. The length of this rectangle along
polar axis is equal to $2\theta_0,$ and along an azimutal one -- $2\Phi.$
Their values are $$\theta_0 = 0.051/16, \qquad\Phi = 3\pi/32 .$$

The cuts, for the cases when a photon and positron form or do not form
cluster CALO2, are very simply formulated in the terms of the positron and
photon variables, namely
\begin{equation}\label{24} |\theta_+ -
     \theta_{\gamma}| < \theta_0 , \ \ 1 >\cos\varphi_{\gamma} >\cos\Phi,
\ \ \ (\gamma, e^+) \in CL ,
\end{equation}
and also
$$|\theta_+ - \theta_{\gamma}| > \theta_0 , \ \ 1 >\cos\varphi_{\gamma}
> -1;$$
\begin{equation}\label{25}
|\theta_0 - \theta_{\gamma}| < \theta_0 , \
\ -1 <\cos\varphi_{\gamma} <\cos\Phi, \ \ \ (\gamma, e^+) \not\in CL ,
\end{equation}
where $\theta_{\gamma} = \vec k^{\bot}/(\varepsilon(1-x)),$ and
$\varphi_{\gamma}$ is the angle between two--dimensional vectors $
\vec k^{\bot}$ and $\vec q_2^{\bot}.$

It is necessary to express the conditions (24) and (25) in terms of
electron and positron variables ${ z,z_1}$ and $\varphi.$ Note in this
case, that the law of the transverse momentum conservation leads to the
following relations:
\begin{equation}\label{26} \cos\varphi_{\gamma} =
- \frac{\theta_-\cos\varphi + x\theta_+}{(1-x)\theta_ {\gamma}},\qquad
\theta_{\gamma} = \frac{1}{1-x}\sqrt{\theta_-^2 + x^2\theta_+^2 +
2x\theta_+\theta_-\cos\varphi} \ .
\end{equation}
Using these relations we take for the case when photon and positron form
a cluster
$$  A < \varphi < B,
\ \ A = max\biggl[-1 =\frac{(1-x)^2(\theta_+ -\theta _0)^2 - (\theta_- -
x\theta_+)^2}{2x\theta_+\theta_-}, \ \ -1\biggr], $$
\begin{equation}\label{27}
B = min\biggl[-1 + \frac{(1-x)^2(\theta_+ +\theta
_0)^2 - (\theta_- - x\theta_+)^2}{2x\theta_+\theta_-}, \ \
\cos(\Phi-\delta)\biggr],\ \ \sin\delta = \sqrt\frac{x\theta_+}{\theta_-}\sin
\Phi.
\end{equation}

From Eq.(27) it follows that there are three different kinematic regions
if photon and positron belong to the cluster CALO2. In the
first region we have $$   -1 +
\frac{(1-x)^2(\sqrt{z_1}-x\overline\lambda)^2
-x^2\sqrt{z}-\sqrt{z_1})^2}{2x^2\sqrt{zz_1}} < \cos\varphi < -\cos(\Phi
-\delta)\ , $$
\begin{equation}\label{28}
 x^2z_+ < z_1 < x^2J_+^2 .
\end{equation}
In the second one
\begin{equation}\label{29}
     -1  < \cos\varphi < -\cos(\Phi-\delta) \ , \quad x^2z_+ > z_1 >
x^2J_-^2 ,
     \end{equation}
and in the third region
\begin{equation}\label{30}
 -1 + \frac{(1-x)^2(\sqrt{z_1}+x\overline\lambda)^2
-x^2\sqrt{z}-\sqrt{z_1})^2}{2x^2\sqrt{zz_1}} > \cos\varphi > -1 \ ,
\quad x^2z_+ < z_1 < x^2J_+^2 \ ,
\end{equation}
where
$$J_{(\pm)} = \frac{1}{\beta}\bigl[\sqrt{z\beta - x^2(1-x)^2\overline\lambda^2
\sin^2\Phi}\  \pm \ (1-x)\overline\lambda(1-2x\sin^2\frac{\Phi}{2})\bigr] \ ,
$$
$$ \beta = 1-4x(1-x)\sin^2\frac{\Phi}{2} \ , \quad \overline\lambda =
\frac{\theta_0}{\theta_1} \ .$$

If the photon and positron do not form the cluster CALO2, then it is
evident that
\begin{equation}\label{31} -1 < \cos\varphi < 1 \ , \quad
     z_1 > x^2J_+^2 \ , \quad z_1 < x^2z_-,
\end{equation}
and in the opposite case
$$  - 1 + \frac{(1-x)^2(\sqrt{z_1}-x\overline\lambda)^2
-x^2\sqrt{z}-\sqrt{z_1})^2}{2x^2\sqrt{zz_1}} > \cos\varphi > -1 \ ,
\quad -\cos(\Phi-\delta) < \cos\varphi < 1,\ \ $$
\begin{equation}\label{32}
\qquad x^2z_+ < z_1 < x^2J_+^2 \ ,
\end{equation}
\begin{equation}\label{33}
-\cos(\Phi-\delta) < \cos\varphi < 1 \ ,\quad x^2z_+ > z_1 > x^2J_-^2 \ ,
\end{equation}
\begin{equation}\label{34}
     -1 + \frac{(1-x)^2(\sqrt{z_1}+x\overline\lambda)^2
-x^2\sqrt{z}-\sqrt{z_1})^2}{2x^2\sqrt{zz_1}} < \cos\varphi <  1 \ ,
\quad x^2z_- < z_1 < x^2J_-^2 \ .
\end{equation}

Integration of Eq.(17) leads to the following result for the contribution
to the part of the first order correction which depends on the shape and
size of the cluster CALO2
$$\Sigma_{1i}^{^{cl}} =
\frac{\alpha}{2\pi}\int\limits_{\ \ \ 0}^{x_c}\frac{1+x^2}
{1-x}dx\int\frac{dz}{z^2}\int dz_1[(\Psi_1 + \widetilde\Psi_1)\Phi_1 +
(\Psi_2 + \widetilde\Psi_2)\Phi_2 +$$
\begin{equation}\label{35} +(\Psi_3
+ \widetilde\Psi_3)\Phi_3]\frac{2}{\pi}\biggl(\frac{1}{z_1-xz} +
\frac{1}{z-z_1}\biggr),
\end{equation}
$$\Sigma_{1f}^{^{cl}} = \frac{\alpha}{2\pi}\int\limits_{\ \ \
0}^{x_c}\frac{1+x^2} {1-x}dx\biggl\{\int\frac{dz}{z^2}\int
dz_1[(\Psi_1 + \widetilde\Psi_1)F_1 + (\Psi_4 +
\widetilde\Psi_4)F_2 +$$ $$ +(\Psi_3 +
\widetilde\Psi_3)F_3]\frac{2}{\pi}\biggl(\frac{1}{z_1-xz} -
\frac{1}{z_1-x^2z}\biggr) +$$
$$ + \int\limits_{\ \ 1}^{\rho_3^2}\frac{dz}{z^{2}}[\theta(max(1,z_2^{(-)}
-z) - \theta(z-min((\rho_4+\overline\lambda(1-x))^2, \rho_3^2)]\widetilde L_4 +$$
$$+ \int\limits_{\ \ \ \tilde a^2}^{\tilde b_0^2}\frac{dz}{z^2}\biggl(\ln\left
|\frac{x\rho_2^2-z}{\rho_2^2-z}\right| + l_-\biggr) +
\int\limits_{\ \ \ max(1,z_2^{(-)})}^{z_4^{(-)}}\frac{dz}{z^2}\ln\left|\frac
{(x\rho_4^2-z)(J_+^2-z)}{(xJ_+^2-z)(\rho_4^2-z)}\right| + $$
\begin{equation}\label{36}
\int\limits_
{ \ \ \rho_2^2}^{min(\rho_4^2,z_3^{(-)})}\frac{dz}{z^2}\ln\left|\frac
{(x\rho_3^2-z)(J_+^2-z)}{(xJ_+^2-z)(\rho_3^2-z)}\right| + \int\limits_
{\ \ \ a^2}^{\rho_4^2}\frac{dz}{z^2}\biggl(\ln\left|\frac{z-x}{1-z}\right|
+ l_-\biggr)\biggr\},
\end{equation}
where
$$  z_i^{(\pm)} = (\rho_i \pm
(1-x)\overline\lambda)^2 - 4x(1-x)\rho_i(\rho_i
\pm\overline\lambda)\sin^2\frac{\Phi}{2 }.$$
The $\Phi_i$ and $F_i$ functions, entering in Eqs.(35) and (36), are
defined as follows
$$\Phi_1 = \arctan Q_i^{(-)} - \arctan \eta, \ \
\Phi_2 = \arctan \eta^{-1}, \ \ \Phi_3 = \arctan\frac{1}{Q_i^{(+)}},$$ $$
F_1 = \arctan\frac{1}{Q_f^{(-)}}, \ \ F_2 = \arctan\zeta, \ \ F_3 =
\arctan \frac{1}{Q_f^{(+)}} ,\ \ \ \ \eta = R_i\cot\frac{\Phi-\delta}{2},
$$
$$\zeta = R_f\cot\frac{\Phi-\delta} {2}, \ \ \ Q_f^{(\pm)} =
\frac{R_f}{R_i}Q_i^{(\pm)},\ \ R_i = \frac{(\sqrt{z}-\sqrt{z_1})^2}{z-z_1},
\ \ R_f = \frac{(\sqrt{z_1}-x\sqrt{z})^2}{z_1-x^2z} \ ,$$
\begin{equation}\label{37} Q_i^{(\pm)} =
     R_i\sqrt{\frac{x^2(\sqrt{z}+\sqrt{z_1})^2 - (1-x)^2( \sqrt{z_1}\pm
     x\overline\lambda)^2}{(1-x)^2(\sqrt{z_1}\pm x\overline\lambda) ^2 -
x^2(\sqrt{z}-\sqrt{z_1})^2}} \ .
\end{equation}
The $\Psi_i$ and $\widetilde\Psi_i$ quantities are inserted in Eqs.(35)
and (36) in order to indicate the integration limits over
${z}$ and ${z_1}$
$$ \Psi_1 =
[min(\rho_4^2, z_3^{(-)})\ , \ \rho_2^2](x^2J_+^2, x^2z_+) + [b^2,
z_3^{(-)}](x^2\rho_3^2, x^2z_+) \ , $$
$$ \Psi_2 = [max(\rho_2^2,
z_1^{(+)})\ ,\ \rho_2^2](x^2z_+, x^2) + [b^2,
max(\rho_2^2,z_1^{(+)}](x^2z_+, x^2J_-^2) + [\rho_4^2, b^2](x^2\rho_3^2 ,
x^2J_-^2) \ ,$$
$$ \Psi_3 = [a^2, max(\rho_2^2, z_1^{(+)})](x^2J_-^2, x^2)
+ [\rho_4^2, a^2](x^2J_-^2, x^2z_-) \ ,$$
$$ \Psi_4 = [max(\rho_2^2,
z_1^{(+)}), \rho_2^2](x^2J_+^2, x^2) + [min(\rho_4^2, z_3^{(-)}),
max(\rho_2^2,z_1^{(+)}](x^2J_+^2, x^2J_-^2) +$$
$$ + [\rho_4^2,
min(\rho_4^2, z_3^{(-)})](x^2\rho_3^2 , x^2J_-^2) \ ,$$
$$\widetilde\Psi_1 =
[(\rho_2^2-\overline\lambda(1-x))^2, max(1,z_2^{(-)}] (x^2J_+^2, x^2\rho_2^2)
+ [z_4^{(-)}, \tilde a_0](x^2J_+^2, x^2z_+) + $$
$$
[(\rho_4-\overline\lambda(1-x))^2, z_4^{(-)})](x^2\rho_4^2, x^2z_+) \ ,$$
$$\widetilde\Psi_2 = [z_2^{(+)}, \tilde a_0^2] (x^2z_+,x^2\rho_2^2) +
[(\rho_4-\overline\lambda(1-x))^2, z_2^{(+)})](x^2z_+, x^2J_-^2) + $$
$$+ [min(\rho_3^2, z_4^{(+)}) \ ,
(\rho_4-\overline\lambda(1-x))^2](x^2\rho_4^2, x^2J_+^2) \ ,$$
$$\widetilde\Psi_3 = [(\rho_2^2+\overline\lambda(1-x))^2, z_2^{(+)}]
(x^2J_-^2, x^2\rho_2^2) + [\tilde b_0, (\rho_2^2+\overline\lambda(1-x))^2]
(x^2J_+^2, x^2z_+)\ ,$$
$$\widetilde\Psi_4 = [z_2^{(+)}, max(1, z_2^{(-)})](x^2J_+^2,x^2\rho_2^2) +
[z_4^{(-)}, z_2^{(+)}](x^2J_+^2, x^2J_-^2) + $$
$$ +[min(\rho_3^2, z_4^{(+)}),
z_4^{(-)}](x^2\rho_4^2, x^2J_-^2) \ .$$
It is necessary to replace $\lambda$ by $\overline\lambda$ in the used
here quantities $ a, b, a_0, b_0.$ As before, in the case of symmetrical
wide detectors, in Eqs.(35) and (36) one must replace $\rho_4$ by $\rho_3$
and $\rho_2$ by 1 and simultaneously put to zero those terms where the
upper integration limit over $z$ becomes less than bottom one.
\vspace{0.3cm}
\begin{center}
\section{Leading second-- and third--order corrections}
\end{center}
\vspace{0.05cm}

As we have seen in the previous Sections, the large logarithm $L$
enters only in the universal part of the radiative correction. Such
situation takes place also for leading contributions in the higher
orders of the perturbation theory. Therefore, the analytical formulae
for high--order leading radiative corrections are universal. They are
applicable for both CALO1 and CALO2 event selections.

In order to obtain the leading contributions in the second and third
orders we use the method based on the electron structure functions in
the singlet and nonsinglet channels [16] and effective coupling constant
[7]
\begin{equation}\label{38}
D(x,\alpha_{eff}) =
D^{NS}(x,\alpha_{eff}) + D^{S}(x,\alpha_{eff}) \ ,
\end{equation}
where the effective coupling constant is defined as the
integral of the running electromagnetic constant in one--loop
approximation
\begin{equation}\label{39}
\frac{\alpha_{eff}}{2\pi} = \int\limits_{\ \
0}^{L}\frac{\alpha dt}{2\pi (1-\frac{\alpha t}{3\pi})} =
\frac{3}{2}\ln\biggl(1-\frac{\alpha L} {3\pi}\biggr)^{-1}.
\end{equation}

The iterative form of the nonsinglet component of the structure function
can be presented as
$$D^{NS}(x,\alpha_{eff}) = \delta(1-x) +
\sum_{k=1}^{\infty}\frac{1}{k!}
\biggl(\frac{\alpha_{eff}}{2\pi}\biggr)^kP_1(x)^{\otimes k} ,$$
\begin{equation}\label{40}
\underbrace{ P_1(x)\otimes\cdots\otimes P_1(x)}_{k} = P_1(x)^{\otimes k} \ ,
\quad P_1(x)\otimes P_1(x) = \int\limits_{\ \
x}^{1}P_1(t)P_1\biggl(\frac{x}{t} \biggr)\frac{dt}{t} \ .
\end{equation}

The singlet component of the electron structure function including
third--order contribution can be written as follows
\begin{equation}\label{41}
D^{S}(x,\alpha_{eff}) =
\frac{1}{2!}\biggl(\frac{\alpha_{eff}}{2\pi}\biggr)^2 R(x) +
\frac{1}{3!}\biggl(\frac{\alpha_{eff}}{2\pi}\biggr)^3\biggl[2P_1\otimes R(x)
- \frac{2}{3}R(x)\biggr] \ ,
\end{equation}
$$ R(x) = \frac{1-x}{3x}(4 + 7x + 4x^2) + 2(1+x)\ln x \ .$$

The nonsinglet component of the electron structure function describes the
photon emission and electron--positron pair production neglecting the
final--electron identity, whereas the singlet one just responds for the
identity effect.

With the required accuracy the electron structure function reads
$$ D(x,L) = \delta(1-x) +
\frac{\alpha L}{2\pi}P_1(x) + \frac{1}{2}\biggl( \frac{\alpha
L}{2\pi}\biggr)^2G(x) + \frac{1}{3}\biggl(\frac{\alpha
L}{2\pi}\biggr)^3F(x), $$
$$G(x) = P_2(x) + \frac{2}{3}P_1(x) + R(x) , \ \
F(x) = \frac{1}{2}P_3(x) + P_2(x) + \frac{4}{9}P_1(x) + \frac{2}{3}R(x) +
R^{^{x}}(x),$$
\begin{equation}\label{42} P_i(x) = P_1^{^{\otimes}i},
\qquad R^{^{^p}}(x) = P_1\otimes R(x) .
\end{equation}
The functions  $P_2(x), P_3(x)$ and $R^{^{p}}(x)$, entering in the right
side of Eq.(42,) can be written as follows [7]
$$R^{^{p}}(x) = [\frac{3}{2}
+ 2\ln(1-x)]R(x) + (1+x)[4L_{i2}(1-x) - \ln ^2x] + \frac{1}{3}(-9 -3x +$$
$$+8x^2)\ln x + \frac{2}{3}(-\frac{3}{x} - 8 + 8x + 3x^2)\ ,$$ $$P_i(x) =
\Delta_i\delta(1-x) + \Theta_i(x)\theta(1-x-\Delta) , \ \ i = 2,\ \ 3 \
,$$
$$\Delta_2 = (2\ln\Delta + \frac{3}{2})^2 - 4\zeta_2 \ ,\ \ \Delta_3 =
(2\ln \Delta + \frac{3}{2})^3 - 12\zeta_2(2\ln\Delta + \frac{3}{2}) +
16\zeta_3 \ ,$$
$$\Theta_2(x) =
2\biggl[\frac{1+x^2}{1-x}(\ln\frac{(1-x)^2}{x} + \frac{3}{2}) +
\frac{1}{2}(1+x)\ln x - 1 +x\biggr],$$ $$\Theta_3(x) =
12\frac{1+x^2}{1-x}\biggl[\ln\frac{1-x}{x}\ln(1-x) + \frac{1} {6}\ln^2x +
\frac{3}{4}\ln\frac{(1-x)^2}{x} + \frac{9}{16} - \zeta_2\biggr] - 3(1-x)
[1+$$
$$+4\ln(1-x)] + \frac{3}{2}(5-3x)\ln x + 6(1+x)\biggl[\frac{1}{4}\ln x
\ln\frac{(1-x)^4}{x} + L_{i2}(1-x)\biggr] .$$
The functions $P_i(x)$ satisfy the condition
$$ \int\limits_{\ \ \ 0}^{1}P_i(x)dx = 0 \ ,$$
which is nothing but the statement of the Lee--Nauenberg theorem
[21] for the leading radiative corrections in the singlet channel in terms
of the electron structure functions.

The factorization form of the differential cross--section in the
frame work of the impact representation [23] determines the leading
contribution to the radiative correction in the case of calorimeter
event selection in the following form
\begin{equation}\label{43} \Sigma^{L} = \int\limits_{\ \ \
0}^{\infty}\frac{dz}{z^2}\int\limits_{\ \ \ x_c}^{1}dx_1
\int\limits_{{x_c}/{x_1}}^{1}dx_2D(x_1,L)D(x_2,L) \Delta_{31}^{(x_1)}
\Delta_{42}^{(x_2)} \ .
\end{equation}
The integrand in Eq.(43) includes only those $\theta$--functions
which depend on ${x_1}$ and ${x_2}.$ It corresponds to account for the
initial--state emission. The final--state emission does not produce the
leading contribution according to the mentioned above Lee--Nauenberg
theorem.

Combining Eqs.(42) and (43) we obtain the leading contribution in the
second and third orders
\begin{equation}\label{44} \Sigma_2^{^{L}} =
     \frac{\alpha^2}{8\pi^2}\int\limits_{\ \ \ 0}^{\infty}
\frac{dz}{z^2}L^2\biggl\{\int\limits_{\ \ \ x_c}^{1}\biggl[(\Delta_{42}^{(x)}
\Delta_{31} + \Delta_{31}^{(x)}\Delta_{42})G(x) + 2\int\limits_{
{x_c}/{x}}^{1}dx_1P_1(x)P_1(x_1)\Delta_{31}^{(x)}\Delta_{42}^{(x_1)}
\biggr]dx\biggr\} \ ,
\end{equation}
 $$  \Sigma_3^{^{L}} = \frac{\alpha^3}{8\pi^3}\int\limits_{\ \ \ 0}^{\infty}
\frac{dz}{z^2}L^3\biggl\{\int\limits_{\ \ \ x_c}^{1}\biggl[
\frac{1}{3}(\Delta_{42}^{(x)}\Delta_{31} + \Delta_{31}^{(x)}\Delta_{42})
F(x) + $$
\begin{equation}\label{45}
+\frac{1}{2}\int\limits_{
{x_c}/{x}}^{1}dx_1P_1(x)G(x_1)(\Delta_{31}^{(x)}\Delta_{42}^{(x_1)} +
\Delta_{31}^{(x_1)}\Delta_{42}^{(x)}\biggr]dx\biggr\} \ .
\end{equation}

In order to eliminate the $\theta$ -- functions, entering in (44) and
(45), and place specific limits, the additional work is needed. Omitting
all intermediate calculations let us write the second--order leading
correction as follows
\begin{equation}\label{46} \Sigma_2^{^{L}} = \Sigma_{2(-)}^{^{L}} +
\Sigma_{2(+)}^{^{L}} + \Sigma_{\gamma}^{^{\gamma}} \ ,
\end{equation}
where the first term on the rihgt side of Eq.(46) is caused due to the
emission of the real and virtual photons as well as electron--positron
pair production by an electron, the second term -- by the positron, and
the third one -- due to the emission of one photon by an electron and one
photon by a positron:
\begin{equation}\label{47}
\Sigma_{2(-)}^{^{L}} =
\frac{\alpha^2}{8\pi^2}\biggl[\int\limits_{\ \ \ \rho_2^2}^{\rho_4^2}\frac{dz}
{z^2}L^2A - \int\limits_{\ \ \ \ m_{23}}^{\rho_4^2}\frac{dz}{z^2}L^2B\biggl(
\frac{\sqrt{z}}{\rho_3}\biggr)\biggr] \ ,
\end{equation}
\begin{equation}\label{48}
\Sigma_{2(+)}^{^{L}} =
\frac{\alpha^2}{8\pi^2}\biggl[\int\limits_{\ \ \ \rho_2^2}^{\rho_4^2}\frac{dz}
{z^2}L^2A - \int\limits_{\ \ \ \ m_{14}}^{\rho_4^2}\frac{dz}{z^2}L^2B\biggl(
\frac{\sqrt{z}}{\rho_4}\biggr)\biggr] + \int\limits_{\ \ \ \
m_{12}}^{\rho_2^2}
\frac{dz}{z^2}L^2B\biggl(\frac{\sqrt{z}}{\rho_2}\biggr)\biggr] \ ,
\end{equation}
$$\Sigma_{\gamma}^{^{\gamma}} =
\frac{\alpha^2}{4\pi^2}\biggl\{-\int\limits_{
\ \ \ \rho_2^2}^{\rho_4^2}\frac{dz} {z^2}L^2F_2(x_c) -
\int\limits_{\ \ \ \ m_{23}}^{\rho_4^2}\frac{dz}{z^2}L^2F_{g}\biggl(
\frac{\sqrt{z}}{\rho_3},x_c\biggr)\biggr] -
\int\limits_{\ \ \ \ m_{14}}^{\rho_4^2}
\frac{dz}{z^2}L^2F_g\biggl(\frac{\sqrt{z}}{\rho_4},x_c\biggr) +$$
$$ + \int\limits_{\ \ \ \ m_{12}}^{\rho_2^2}
\frac{dz}{z^2}L^2F_g\biggl(\frac{\sqrt{z}}{\rho_2},x_c\biggr) +
\int\limits_{\ \ x_c\rho_3\rho_4}^{\rho_4^2}\frac{dz}{z^2}L^2\biggl[
F_g\biggl(\frac{\sqrt{z}}{\rho_4},\frac{x_c\rho_3}{\sqrt{z}}\biggr)
+ C\biggl(\frac{\sqrt{z}}{\rho_3},\frac{\sqrt{z}}{\rho_4}\biggr)\biggr] + $$
$$\int\limits_{\ x_c\rho_2}^{1}\frac{dz}{z^2}L^2\biggl[
F_g\biggl(\sqrt{z},\frac{x_c\rho_2}{\sqrt{z}}\biggr)
+ C\biggl(\frac{\sqrt{z}}{\rho_2},\sqrt{z}\biggr)\biggr] -
\int\limits_{\ x_c\rho_4}^{1}\frac{dz}{z^2}L^2\biggl[
F_g\biggl(\frac{\sqrt{z}}{\rho_4},\frac{x_c}{\sqrt{z}}\biggr) + $$
\begin{equation}\label{49}
+ C\biggl(\sqrt{z},\frac{\sqrt{z}}{\rho_4}\biggr)\biggr] -
\int\limits_{\ \ x_c\rho_3\rho_2}^{\rho_2^2}\frac{dz}{z^2}L^2\biggl[
F_g\biggl(\frac{\sqrt{z}}{\rho_3},\frac{x_c\rho_2}{\sqrt{z}}\biggr)
+ C\biggl(\frac{\sqrt{z}}{\rho_2},\frac{\sqrt{z}}{\rho_3}\biggr)\biggr]
\biggr\},
\end{equation}
where
$$ m_{ij} = max(x_c^2\rho_j^2, \rho_i^2) \ , \
A = -\int\limits_{\ \ \ 0}^{x_c}(P_2(x) + \frac{2}{3}P_1(x))dx + \int\limits
_{\ \ \ x_c}^{1}R(x)dx, \ \ \ B(y) = \int\limits_{\ \ \ x_c}^{y}G(x)dx ,$$
$$F_2(x) = \int\limits_{\ \ \ 0}^{x}P_2(y)dy = -2x - \frac{x^2}{4} +
(x+\frac{x^2} {2})\ln\frac{x^3}{(1-x)^4} + 4\ln(1-x)\ln\frac{x}{1-x} +
4L_{i2}(x) \ , $$ $$F_g(x,y) = F_g(x) - F_g(y) ,$$ $$F_g(x) = \int
P_1(x)g\left(\frac{x_c}{x}\right)dx = - \frac{x_c^2}{x} + (2x + x^2)\ln x +
\left(x_c+\frac{x_c^2}{2}\right)\ln\frac{x}{(1-x)^2} + \biggl( 2x_c +
\frac{x_c^2}{2}-$$ $$-2x-\frac{x^2}{2}\biggr)\ln(1-x_c) + 4L_{i2}(x) +
4L_{i2}\left(\frac{1-x}{1-x_c}\right) , \ \ x_c < x < 1 \ , $$ $$ C(x,y) =
g(x)\left[g(y) - g\left(\frac{x_c}{x}\right)\right] \ , $$
\begin{equation}\label{50}
g(x) = - \int P_1(x)dx = x + \frac{x^2}{2} + 2\ln(1-x) , \ \ x < 1 \ .
\end{equation}

By analogy with Eq.(46) the third--order leading correction can be
written as follows
\begin{equation}\label{51} \Sigma_3^{^{L}} =
\Sigma_0^{^{3}} + \Sigma_3^{^{0}} + \Sigma_1^{^{2}} + \Sigma_2^{^{1}} \ ,
\end{equation}
where the upper index shows the number of real and virtual particles
emitted by an electron and the bottom one -- by a positron. For the case
when three additional particles are emitted by one of the fermions, the
corresponding contribution to $\Sigma_3^{^{L}}$ reads
$$\Sigma_0^{^{3}} + \Sigma_3^{^{0}} = \left(\frac{\alpha}{2\pi}\right)^3
\biggl\{\int\limits_{\ \ \
\rho_2^2}^{\rho_4^2}\frac{dz}{z^2}L^{^{3}}\biggl[ -2 \int\limits_{\ \ \
0}^{x_c}F_p(x)dx + 2\int\limits_{\ \ \ x_c}^{1}F_r(x)dx\biggr] - $$
\begin{equation}\label{52} - \int\limits_{\ \ \ \
m_{23}}^{\rho_4^2}\frac{dz}{z^2}L^{^{3}}\int\limits_{\ \ \ x_c}^
{\sqrt{z}/\rho_3}F_c(x)dx - \int\limits_{\ \ \
\ m_{14}}^{\rho_4^2}\frac{dz}{z^2}
L^{^{3}}\int\limits_{\ \ \ x_c}^{\sqrt{z}/\rho_4}F_c(x)dx +
\int\limits_{\ \ \ \ m_{12}}
^{\rho_2^2}\frac{dz}{z^2}L^{^{3}}\int\limits_{\ \
\ \ x_c}^{\sqrt{z}/\rho_2}F_c(x) dx\biggr\} \ , \end{equation} £¤¥ $$F_p(x) =
\frac{1}{6} P_3(x) + \frac{1}{3}P_2(x) + \frac{4}{27}P_1(x) \ , $$ $$F_r(x) =
\frac{2}{9}R(x) + \frac{1}{3}R^{^{p}}(x)\ , \quad F_c(x) = \frac{1}{3}F(x) \
. $$
In the case when both fermions radiate simultaneously, we have
$$\Sigma_1^{^{2}} + \Sigma_2^{^{1}} = \left(\frac{\alpha}{2\pi}\right)^3
\biggl\{\int\limits_{\ \ \ \rho_2^2}^{\rho_4^2}\frac{dz}{z^2}L^{^{3}}\biggl
[-\int\limits_{\ \ \ 0}^{x_c}\left(P_3(x) + \frac{2}{3}P_2(x)\right)dx + \int
\limits_{\ \ \  x_c}^{1}R^{^{p}}(x)dx\biggr] - $$
$$- \int\limits_{\ \
\ \ m_{23}}^{\rho_4^2}\frac{dz}{z^2}L^{^{3}}\int\limits_{\ \ \ x_c}
^{\sqrt{z}/\rho_3}H(x,x_c)dx - \int\limits_{\
\ \ \ m_{14}}^{\rho_4^2}\frac{dz}{z^2}
L^{^{3}}\int\limits_{\ \ \ x_c}^{\sqrt{z}/\rho_4}H(x,x_c)dx +
\int\limits_{\ \ \ \ m_{12}}
^{\rho_2^2}\frac{dz}{z^2}L^{^{3}}\int\limits_{\
\ \ x_c}^{\sqrt{z}/\rho_2}H(x,x_c)dx + $$ $$+
\int\limits_{\ \ x_c\rho_3\rho_4}^{\rho_4^2}\frac{dz}{z^2}L^{^{3}}\biggl[\int
\limits_{x_c\rho_4/\sqrt{z}}^{\sqrt{z}/\rho_3}dxP_1(x)N\left(\frac{x_c}{x};\frac
{\sqrt{z}}{\rho_4}\right)dx + (\rho_3 \leftrightarrow\rho_4)\biggr] +$$
$$+ \int\limits_{\ x_c\rho_2}^{1}\frac{dz}{z^2}L^{^{3}}\biggl[\int\limits_{
x_c\rho_2/\sqrt{z}}^{\sqrt{z}/1}dxP_1(x)N\left(\frac{x_c}{x};\frac{\sqrt{z}}
{\rho_2}\right)dx + (\rho_2\leftrightarrow 1)\biggr] - $$
$$- \int\limits_{\ \ x_c\rho_2\rho_3}^{\rho_2^2}L^{^{3}}\biggl[\int\limits_{
x_c\rho_2/\sqrt{z}}^{\sqrt{z}/\rho_3}dxP_1(x)N\left(\frac{x_c}{x};\frac{\sqrt
{z}}{\rho_2}\right)dx + (\rho_3\leftrightarrow\rho_2)\biggr] - $$
\begin{equation}\label{53}
- \int\limits_{\ x_c\rho_4}^{1}\frac{dz}{z^2}L^{^{3}}\biggl[\int\limits_{
x_c\rho_4/\sqrt{z}}^{\sqrt{z}/1}dxP_1(x)N\left(\frac{x_c}{x};\frac{\sqrt{z}}
{\rho_4}\right)dx + (\rho_4\leftrightarrow 1)\biggr]\biggr\},
\end{equation}
where
$$H(x,x_c) = P_1(x)\left[\frac{1}{2}f\left(\frac{x_c}{x}\right) + \frac{2}{3}
g\left(\frac{x_c}{x}\right) + \frac{1}{2}r\left(\frac{x_c}{x}\right)\right]
+\frac{1}{2}g\left(\frac{x_c}{x}\right)(P_2(x) + R(x)), $$
$$N(x;y) = N(x) - N(y), \qquad N(y) = \frac{1}{2}f(y) + \frac{1}{3}g(y) +
\frac{1}{2}r(y), $$
$$f(y) = -F_2(y),\ \ r(y) = \int\limits_{\ \ y}^{1}R(x)dx = -\frac{22}{9} + y
+y^2 + \frac{4}{9}y^3 - \left(\frac{4}{3}+2y+y^2\right)\ln y .$$

When writing the formulae of this Section, we represent the restrictions
on the angles and energies of the detected particles with the help of
definite integrals using such relations as, for example,
\begin{equation}\label{54}
\int\theta_4\overline\theta_4^{(x)}\overline\theta_3^{(x_1)}dz\,dx\,dx_1 =
\int\limits_{x_c\rho_3}^{\rho_4^2}dz\int\limits_{x_c\rho_3/\sqrt{z}}^{\sqrt{z}
/\rho_4}dx\int\limits_{x_c/x}^{\sqrt{z}/\rho_3}dx_1 \ .
\end{equation}
It is necessary to bear in mind that the upper limit of the integration
over the variable $z$ always must be greater than bottom one.
In opposite case the integral must be putted to zero.
\vspace{0.3cm}
\begin{center}
\section{Numerical results}
\end{center}
\vspace{0.05cm}

At LEP1 conditions the limiting angles of the ring--shaped detectors have
different values for various versions of event selection. In the cases of
BARE1 and CALO1 the wide detector has
$$\theta_1 = 0.024 \ , \qquad
\theta_3 = 0.058 \ ,$$
and narrow one --
$$\Theta_2 = \theta_1+h \ ,\ \ \theta_4 =
\theta_3-h\ ,\ \ h = \frac{0.017}{8}\ .$$
The wide detector ({\bf ww}) for the event selection CALO2 coinsides with
the narrow one ({\bf nn}) for the case BARE1, whereas the limiting angles
of the narrow detector are determined as follows
$$\theta_2
= \theta_1 + 2h \ , \quad \theta_4 = \theta_3 - 4h \ .$$

The Born cross--section, determined by the formula (4), is equal to

$\sigma_B$ = 175.922 {\bf nb} for {\bf ww} BARE1 ¨ CALO1,

$\sigma_B$ = 139.971 {\bf nb} for {\bf ww} CALO2, {\bf nn} BARE1 and
CALO2,

$ \sigma_B$ = 103.299 {\bf nb} ¤«ï {\bf nn} CALO2.

The results of our calculations of the radiative QED corrections are
presented in the Tables I--III, where the vacuum polarization is switched
off. For the comparison we also present numbers in the Tables I and III,
that are obtained with the help of the MC--generator BHLUMI using the
exponentiated Yennie--Frautchi--Suura factor for the higher--order
corrections.

\vspace{0.5cm}
\begin{center}
\begin{tabular}{|ccccc|}  \hline $x_c$   & {\bf bhlumi ww}  & {\bf ww}  & {\bf
nn}  & {\bf wn} \\ \hline \multicolumn{5}{|c|}{\bf calo1 } \\  \hline 0.1 &
166.329 & 166.285 & 131.032 & 134.270 \\ 0.3 & 166.049 & 166.006 & 130.833
& 134.036 \\ 0.5 & 165.287 & 165.244 & 130.416 & 133.466 \\ 0.7 & 161.794 &
161.749 & 128.044 & 130.542 \\ 0.9 & 149.925 & 149.866 & 118.822 & 120.038
\\ \hline \multicolumn{5}{|c|}{\bf calo2 } \\ \hline 0.1 & 131.032 & 130.997
& 94.666  & 98.354   \\ 0.3 & 130.739 & 130.705 & 94.491  & 98.127   \\ 0.5
& 130.176 & 130.141 & 94.177  & 97.720   \\ 0.7 & 127.528 & 127.491 & 92.981
& 95.874   \\ 0.9 & 117.541 & 117.491 & 86.303  & 87.696   \\  \hline
\end{tabular}
\end{center}
\hspace{4.2cm}{\small{\bf Table I}.{SABH cross--section at LEP1
conditions with \\ }
\hspace*{4.9cm}{first-- order correction. Vacuum polarization  \\ }
\hspace*{6.8cm}{is switched off \\ }}
\vspace{0.5cm}

From the Table I one can see, that at LEP1 conditions, the SABH
cross--section including the first--order correction, obtained by the
means of BHLUMI generator systematically exceeds our results in all
interval of values for parameter $x_c$. The relative difference amounts to
about 0.03\%. The possible reason of this defference, as it was noted in
Section 2, is that we neglect the contributions proportional
to $\theta_1^2$ when calculating the radiative corrections. Within the
accuracy of 0.1\% the contribution of the omitted by us terms is
insignificant. But it requires the additional systematic investigations if
the accuracy will be  0.05\% (and it had actually been achieved in the
laboratory L3 [24]).

The absolute values of the second-- and third--order leading corrections
(in ${\bf {nb}}$ are presented in the Table II. The second--order
correction is divided into the contributions related to the
electron--positron pair production and double--photon emission. One can
see that photonic corrections dominate in the second order of the
perturbation theory. The third--order correction includes the
contributions due to three--photon emission and pair production
accompanied by single--photon emission.

\vspace{0.5cm}
\begin{center}
\begin{tabular}{|ccccccc|} \hline \multicolumn{7}{|c|}{\bf
calo1 \hspace{4cm} calo2} \\  \hline $x_c$ & {\bf ww} & {\bf nn} & {\bf wn} &
{\bf ww} & {\bf nn} & {\bf wn} \\ \hline \multicolumn{7}{|c|}{\bf
correction due to pair productiion} \\ \hline 0.1
&--0.046&--0.045&--0.024&--0.045&--0.047&--0.024 \\ 0.3
&--0.046&--0.045&--0.024&--0.045&--0.047&--0.024 \\ 0.5
&--0.048&--0.046&--0.025&--0.046&--0.047&--0.024 \\ 0.7
&--0.069&--0.059&--0.042&--0.059&--0.051&--0.036 \\ 0.9
&--0.137&--0.111&--0.102&--0.111&--0.085&--0.075 \\ \hline
\multicolumn{7}{|c|}{\bf second order photonic correction} \\  \hline
0.1 & 0.788& 0.708& 0.302& 0.708& 0.668& 0.255 \\
0.3 & 0.680& 0.634& 0.195& 0.634& 0.627& 0.187 \\
0.5 & 0.474& 0.487& 0.005& 0.487& 0.554& 0.073 \\
0.7 & 0.293& 0.317& --0.164& 0.317& 0.396& --0.092 \\
0.9 & 0.866& 0.738& 0.373& 0.738& 0.628& 0.271 \\ \hline
\multicolumn{7}{|c|}{\bf third order correction} \\ \hline
0.1 & --0.041& --0.036&--0.002& --0.036&--0.034&--0.001 \\
0.3 & --0.046& --0.040&--0.007& --0.040&--0.037&--0.003 \\
0.5 & --0.044& --0.039&--0.006& --0.039&--0.037&--0.005 \\
0.7 & --0.023& --0.022&  0.012& --0.022&--0.027&  0.008 \\
0.9 &   0.021&   0.013&  0.049&   0.013&  0.002&  0.038 \\ \hline
\end{tabular}
\end{center}
\hspace{4.3cm}{\small {\bf Table II} {Absolute values of second-- and
third--order \\ }
\hspace*{3.6cm} {corrections to SABH cross--section at
LEP1 conditions (in {\bf nb)} \\}} \vspace{0.5cm}

In the Table III we present the total SABH cross--section at LEP1 with
account for all corrections calculated in this paper and also the results
of the corresponding calculations made with the help of the MC--generator
BHLUMI.

As to comparison of our calculations and MC--generator BHLUMI ones for the
radiative corrections in the second and third orders, it is necessary to
note that BHLUMI results are based on the exponentiated form of electron
structure function whereas our ones -- on the iterative form. The
corresponding effect due to these different forms increases at large
values of parameter $x_c$ as one can see from the Table III.

For more efficient comparison, it is necessary to have either the
analytical calculations with the exponential form of electron structure
functions or MC--calculations that do not use the exponentiation. For MC
generator BHLUMI such calculations exist in the case of BARE1 event
selection [3]. The corresponding comparison have been recently done and
the agreement has been obtained at the very high level [25].

\vspace{0.5cm}
\begin{center}
\begin{tabular}{|ccccccc|}  \hline
\multicolumn{7}{|c|}{\bf calo1 \hspace{5.0cm} calo2} \\ \hline
$x_c$ &{\bf ww}&{\bf nn}& {\bf wn}& {\bf ww}& {\bf nn}& {\bf wn} \\ \hline
\multicolumn{7}{|c|}{\bf total cross--section (analytical
calculation)} \\ \hline 0.1 &
166.968&131.659&134.546&131.624&95.253&98.584 \\ 0.3 &
166.594&131.382&134.200&131.254&95.032&98.285 \\ 0.5 &
165.626&130.818&133.440&130.543&94.647&97.764 \\ 0.7 &
161.950&128.280&130.348&127.727&93.299&95.758 \\ 0.9 &
150.616&119.462&120.358&118.131&86.848&87.930 \\ \hline
\multicolumn{7}{|c|}{\bf total cross--section (bhlumi--generator)}  \\
\hline 0.1& 167.203&  &  & 131.835&95.458&98.834 \\ 0.3& 166.795&  &  &
131.450&95.233&98.539 \\ 0.5& 165.830&  &  & 130.727&94.841&98.020 \\ 0.7&
162.237&  &  & 127.969&93.520&96.054 \\ 0.9& 151.270&  &  &
118.792&87.359&88.554  \\ \hline \end{tabular} \end{center}
\hspace{3cm}{\small {\bf Table III} {Total SABH  cross--section at LEP1
conditions \\ }}
\vspace{0.5cm}

The authors are grateful to A.Arbuzov, E.Kuraev and L.Trentadue for
fruitful discussions and critical remarks. One of us (N.P.M.)would like to
thank S.Jadach, B.Ward, G.Montagna and B.Pietrzyk for discussion of
the results of this paper. This work was supported in part by INTAS
Grant 93--1867 ext.

\begin{center}
{ References \\}
\end{center} \begin{enumerate}
\item G. Barbiellini et al. L. Trentadue (conv.), {\it Neutrino
Counting in Z Physics at LEP};  G. Altarelli, R. Kleiss, G. Verzegnassi,
 {\it CERN Yellow Report 89--08} ;  {\it CERN Yellow Report 95--03} \
 , Part I \ : {\it "Electroweak Physics"}.
\item {B. Pietrzyk,  \
{\it Preprint LAPP--EXP--94.18.} Invited Talk on International Symposium
on Radiative Corrections: Status and Outlook, \ Galtinburg, Tennessee,
USA, June 1994, edited by Ward B.W.L. (World Scientific, Singapore,
1995) \ ; \\ } LEP Electroweak Working Group.\ {\it CERN Report
LEPEWWG/95--02} \ ; LEP Collaborations, 1995, Collaboration notes:  ALEPH
95--093 PHYSICS 95--086 \ ; \ \ DELPHI 95--137 PHYS 562 \ ; \ \ L3 Note
1814 \ ;\ \ OPAL Technical Note TN 312 \ , 1 August 1995.
\item {H. Anlauf et al., Events Generator for Bhabha Scattering. In \
:  Conveners: S. Jadach, O. Nicrosini}.  {\it CERN Yellow Report 96--01},
V.2. P.229. \ ;  {\it CERN Yellow Report 95--03} \ , Part III: "Small
Angle Bhabha Scattering".
\item S. Jadach, E. Richter-Was E., B.F.L. Ward, Z. Was,
{\it Comput. Phys.  Commun.} {\bf 70} (1992) 305.
\item
G. Montagna et al., {\it Comput.  Phys. Commun.} {\bf 76} (1993) 328 .  \\
M. Cassiori, G. Montagna, F. Piccinini, {\it Comput.  Phys.  Commun.}
{\bf 90} (1995) 301. \\  G. Montagna et al., {\it Nucl. Phys.} {bf B
401} (1993) 3.
\item S. Jadach, E. Richter-Was, B.F.L. Ward, Z. Was,
 {\it Phys. Lett.} {\bf B 353} (1995); 349, 362 \ . \\
S. Jadach, M. Melles, B.F.L. Ward, S.A. Yost, {\it Phys. Lett.} {\bf
B 377} (1996) 168.
\item S. Jadach M. Skrzypek, B.F.L. Ward,
{\it Phys.  Rev.} {\bf D 47} (1993) 3733 \ .  \\ S. Jadach,
E. Richter--Was, B.F.L. Ward, Z. Was,  {\it Phys. Lett.} {\bf B 260}
(1991) 438.
\item W. Beenakker, F.A. Berends, S.C. van der Marck,
{\it Nucl. Phys.} {\bf B 355} (1991) 281. \\ W. Beenakker,
B. Pietrzyk, {\it Phys. Lett.} {\bf B 304} (1993) 366.
\item M. Gaffo, H. Czyz, E. Remiddi, {\it Nuovo Cim.} {\bf A 105} (1992)
271 ; {\it Int. J. Mod. Phys.} {\bf 4} (1993) 591; {\it Phys.
Lett.} {\bf B 327} (1994) 369.
\item G. Montagna,
O. Nicrosini, F. Piccinini,  {\it Preprint FNT/T--96/8 } \ .
\item A.B. Arbuzov et al., {\it CERN Yellow Report 95--03} 369.
\item A.B. Arbuzov et al., {\it Nucl. Phys.} {\bf B 485} (1997) 457.
\item  A.B. Arbuzov, E.A. Kuraev, N.P. Merenkov, L. Trentadue,
{\it Zh.Eksp.Teor.Fiz.} {\bf 108} (1995) 1164; Preprint CERN--TH/95--241\
, JINR--E2--95--110.
\item N.P. Merenkov et al., {\it Acta Phys.
Pol.} {\bf B 28} (1997) 491.  \\  N.P. Merenkov, {\it JETP}
{\bf 85} (1997) 217; \\ N.P. Merenkov, {\it JETP Lett.} {\bf 65} (1997)
235.
\item
D.R. Yennie, S.C. Frautchi, H. Suura, {\it Ann. Phys.} {\bf 13} (1961)
379.
\item L.N. Lipatov, {\it Sov. Journ. Nucl. Phys.} {\bf 20}
(1974) 121 . \\  G. Altarelli, G. Parisi, {\it Nucl. Phys.} {\bf B 126}
(1977) 298 .\\  M. Skrzypek, {\it Acta Phys. Pol.} {\bf B 23} (1992)
135.
\item M. Gell--Mann, F. Low, {\it Phys. Rev.} {\bf 95} (1954)
1300.
\item V.G. Gorshkov, {\it Usp. Fiz. Nauk} {\bf 110} (1973) 45 .
\\  V.N. Baier, S.A. Kheifez, {\it Zh. Eksp. Teor. Fiz} {\bf 40} (1961)
613 .  \\  V.G. Gorshkov, V.N. Gribov, G.V. Frolov,
{\it Zh. Eksp. Teor. Fiz.} {\bf 51} (1966) 1093 .  \\ V.G. Gorshkov,
V.N. Gribov, L.N. Lipatov, V.G. Frolov, {\it Yad.
Fiz.} {\bf 6} (1967) 129, 361.
\item F.A. Berends
et al., {\it Nucl. Phys.} {\bf B 57} (1973) 371. \\
E.A. Kuraev, G.V. Meledin, {\it Nucl. Phys.} {\bf B 122} (1977)
485.
\item  S. Jadach, B.F.L. Ward, {\it Phys. Rev.} {\bf D 40} (1989)
3582.
\item T.D. Lee, M. Nauenberg, {\it Phys. Rev.} {\bf B 133} (1964)
1549.
\item N.P. Merenkov,  {\it Yad. Fiz.} {\bf 48}
(1988) 1782: {\bf 50} (1989) 750.
\item H. Cheng, T.T. Wu, {\it Phys.
Rev. Lett.} {\bf 23} (1969) 670 .  \\  V.G. Zima,
N.P. Merenkov, {\it Yad. Fiz.} {\bf 25} (1976) 998.
  \\  V.N. Baier, V.S. Fadin, V. Khoze, E.A. Kuraev, {\it Phys. Rep.}
{\bf 78} (1981) 294.
\item I.C. Brock et al.,
{\it Preprint CERN--PPE/96--89' CMU--HEP/96--04.}
\item A.B. Arbuzov et al.,
{\it Phys. Lett} {\bf B 399} (1997) 324.

\end{enumerate}

\end{document}